# X-ray Scaling Relation in Early-Type Galaxies: Dark Matter as a Primary Factor in Retaining Hot Gas

**Dong-Woo Kim & Giuseppina Fabbiano**

Smithsonian Astrophysical Observatory,
60 Garden Street, Cambridge, MA 02138, USA

(August 20, 2013)

**abstract**

We have revisited the X-ray scaling relations of early type galaxies (ETG) by investigating, for the first time, the $L_{X,Gas}$ - $M_{Total}$ relation in a sample of 14 ETGs. In contrast to the large scatter (a factor of $10^2$-$10^3$) in the $L_{X,Total}$ - $L_B$ relation, we found a tight correlation between these physically motivated quantities with a rms deviation of a factor of 3 in $L_{X,Gas}$ = $10^{38}$ - $10^{43}$ erg s$^{-1}$ or $M_{Total}$ = a few x $10^{10}$ - a few x $10^{12}$ $M_\odot$. More striking, this relation becomes even tighter with a rms deviation of a factor of 1.3 among the gas-rich galaxies (with $L_{X,Gas}$ > $10^{40}$ erg s$^{-1}$). In a simple power-law form, the new relation is $(L_{X,Gas} / 10^{40}$ erg s$^{-1}) = (M_{Total} / 3.2$ x $10^{11}$ $M\odot)^3$. This relation is also consistent with the steep relation between the gas luminosity and temperature, $L_{X,Gas} \sim T_{Gas}^{4.5}$, identified by Boroson, Kim & Fabbiano (2011), if the gas is virialized. Our results indicate that the total mass of an ETG is the primary factor in regulating the amount of hot gas. Among the gas-poor galaxies (with $L_{X,Gas}$ < a few x $10^{39}$ erg s$^{-1}$), the scatter in the $L_{X,Gas}$ – $M_{Total}$ (and $L_{X,Gas}$ – $T_{Gas}$) relation increases, suggesting that secondary factors (e.g., rotation, flattening, star formation history, cold gas, environment etc.) may become important.

1. INTRODUCTION

The '$L_X$-$L_B$' relation (or $L_X$-$L_{Optical}$, $L_X$-$L_K$; we will use $L_X$-$L_K$ in this paper) of early-type galaxies (ETG, i.e. E and S0,) has been widely discussed since its first formulation by Trinchieri & Fabbiano (1985) and Forman, Jones & Tucker (1985). This relation, linking the stellar mass of ETGs with their total (~0.5-5 keV) X-ray luminosity, which is an approximated proxy for the amount of hot gas they can retain, has been related to the gravitational confinement of the hot gas, and has been used to investigate the origin and evolution of the hot interstellar medium (ISM) of ETGs, including the effects of interactions in galaxy clusters, and winds from stellar and AGN feedback (e.g., see review in Fabbiano 1989 and Mathews & Brighenti 2003; also Canizares, Fabbiano & Trinchieri 1987; Ciotti et al. 1991; David et al. 1991; White & Sarazin 1991). The structural parameters of the ETG also appear to be relevant for the hot gas retention. 'Boxy' galaxies with a central core tend to have larger amounts of hot gas than 'disky' galaxies with central stellar cusps and fast rotation (Bender et al.1991; Eskridge et al. 1995b; Pellegrini 2005). Core ETG also tend to be larger and with older stellar populations, leading Kormendy et al. (2009) to suggest that the hot ISM may provide the working surface, necessary for feedback, by storing and smoothing episodic energy input and by shielding against the accretion of fresh gas, and thus impeding star formation (see also Binney 2004, Nipoti & Binney 2007).

The main characteristic of the $L_X$-$L_K$ relation, which all the above studies have tried to explain, is its large spread. For a given ETG optical luminosity $L_K$ (i.e., stellar mass), $L_X$ can vary by a factor of ~100 in different ETGs (e.g., Fabbiano 1989; Eskridge et al. 1995a; Ellis et al 2006). However, the real spread is even larger, because the X-ray luminosity $L_X$ used in most published studies, as a proxy of the hot gas content, is the total integrated $L_X$ of an ETG galaxy, which contains a significant contribution from the integrated input of stellar sources, including low-mass X-ray binaries (LMXBs; as first pointed out by Trinchieri & Fabbiano 1985). With the sub-arcsecond resolution of the *Chandra X-ray Observatory*, we are now able to measure the hot gas luminosity, $L_{X,Gas}$, by excluding individually detected LMXBs (see a review by Fabbiano 2006) and nuclear sources. We can also estimate and subtract the contribution of undetected stellar sources (faint LMXBs, active binaries and cataclysmic variables,) by fitting the X-ray spectra with multiple components (Boroson, Kim & Fabbiano 2011 - hereafter BKF) and extrapolating to the low luminosities the X-ray luminosity function of LMXBs (e.g., Fig 4 in Kim & Fabbiano 2010). The accurate subtraction of these contaminants is most critical in gas poor galaxies where the X-ray luminosity of the hot gas can be commensurable with or lower than that of the stellar sources. When the data are cleaned we obtain a range of scatter in $L_{X,Gas}/L_K$ that can be as large as a factor of ~$10^3$ or more (BKF).

This noisy scaling relation suggests that we may be trying to correlate the wrong quantities. The optical luminosity ($L_K$) is a good proxy for the integrated stellar mass of the galaxy, $M_\star$; however, it does not measure the amount of Dark Matter (DM) mass, which may be prevalent, especially at large radii. The total mass (stellar + DM), out to radii comparable to the total extent of the hot halos of gas-rich ETGs is the physical quantity we need to know, if we want to explore the importance of gravitational confinement for the hot gas retention (see Mathews et al. 2006). The amount of gas mass itself is small in ETGs and not important for gravitational confinement (e.g., Canizares, et al. 1987). While dynamical masses have been measured using integral field 2D spectroscopic data for a large number of ETGs (e.g., in the SAURON sample; Cappellari et al. 2010), these data are limited to radii within r < 0.5 - 1 $R_e$ (effective radius or half light radius), smaller than the extent of the hot gas in gas-rich ETGs, and so are not optimal for our purpose. However, recently a number of dynamical mass measurements at large radii have become available, from the analysis of the kinematics of hundreds of globular clusters (GC) and planetary nebulae (PN) in individual galaxies (Deason et al 2012**).

Given these improvements in both X-ray and mass measurements, we have revisited the scaling relations of ETGs by investigating, for the first time, the $L_{X,Gas}$ - $M_{Total}$ relation in a sample of 14 ETGs, for which both X-ray and kinematics data are available. The results are presented in this paper. In Section 2, we describe our sample selection, *Chandra* observations and data reduction techniques. In Section 3, we present the X-ray scaling relations of ETGs. In Section 4, we discuss the implications of our results.

## 2. GALAXY SAMPLE AND X-RAY DATA ANALYSIS

For the total mass of galaxies, we use the direct mass measurements from optical kinematics data of globular clusters (GC) from the SLUGGS survey and planetary nebulae (PN) from the PN.S survey. Deason et al. (2012) compiled optical data for 15 ETGs from the literature and provided a homogeneous dataset of masses within 5 $R_e$. Fourteen of them were observed by Chandra for longer than 15 ksec and are used in this study (see Table 1). NGC 1344 is not used here because its exposure is too shallow (3 ksec) to measure the necessary gas properties.

```
                            Table 1. Galaxy Sample
-------------------------------------------------------------------------------
  name     T      d       log L_K    M(< 5 Re)         L_X,Gas           reference
              (Mpc)     (Lo)      (10^11 Mo)       (10^40 erg/s)
  (1)    (2)    (3)      (4)         (5)               (6)                (7)
-------------------------------------------------------------------------------
  N0821   -5   24.10    10.93     2.7   (0.6)     0.025 (-0.020 +0.022)  this work
  N1399   -5   19.95    11.40    12.8   (1.8)    49.2   (-1.28  +1.28 )  O'Sullivan
  N1407   -5   28.84    11.57    11.0   (1.5)    15.9   (-1.86  +1.86 )  O'Sullivan
  N3377   -5   11.22    10.45     0.7   (0.2)     0.010 (-0.006 +0.007)  this work
  N3379   -5   10.57    10.87     1.4   (0.2)     0.042 (-0.016 +0.016)  this work

  N4374   -5   18.37    11.37    16.5   (2.0)     6.65  (-1.18  +1.18 )  O'Sullivan
  N4486   -4   16.07    11.41    30.6   (3.1)   905.5   (-1.32  +1.32 )  O'Sullivan
  N4494   -5   17.06    10.99     1.2   (0.2)     0.097 (-0.077 +0.078)  this work
  N4564   -5   15.00    10.50     0.4   (0.1)     0.038 (-0.019 +0.020)  this work
  N4636   -5   14.66    11.09    10.7   (1.9)    31.7   (-0.62  +0.62 )  O'Sullivan

  N4649   -5   16.83    11.48     8.7   (1.3)    18.3   (-1.54  +1.54 )  O'Sullivan
  N4697   -5   11.75    10.92     1.5   (0.2)     0.184 (-0.019 +0.038)  this work
  N5128   -2    4.21    11.00     4.9   (0.5)     1.93  (-0.14  +0.14 )  Kraft
  N5846   -5   24.89    11.34    11.7   (2.8)    50.5   (-1.10  +1.10 )  O'Sullivan
-------------------------------------------------------------------------------
(1) galaxy name
(2) morphological type from RC3
(3) distance from Tonry et al. (2001)
(4) K-band luminosity from 2MASS (assuming K_⊙ = 3.33 mag)
(5) total mass within 5 effective radii taken from Deason et al. (2012) after
    correcting for slightly different distances
(6) X-ray luminosity in 0.3-8 keV from the hot gas (see reference), with the error as
    explained in the text
(7) references for L_X,Gas
    O'Sullivan et al. (2001)
    Kraft et al. (2003)
    this work (see Table 2)
```

For six gas-poor galaxies, we measure $L_{X,Gas}$ with the archived *Chandra* ACIS data (see Table 2). All galaxies were observed on the back-illuminated S3 chip, which has the better response at the energy range of interest (< 2 keV) than the front-illuminated chip. As the entire gas emission is typically confined within a single CCD chip, we only analyze the data on the S3 chip. In hot gas-poor galaxies, the X-ray luminosity of hot gas, $L_{X,Gas}$, is often lower than that of LMXBs and in extremely gas-poor galaxies, $L_{X,Gas}$ is even lower than that of active binaries (AB) and cataclysmic variables (CV), $L_{X,AB+CV}$, which is about 1/10 of $L_{X,LMXB}$ (see BKF). Therefore, to accurately determine $L_{X,Gas}$ it is critical to properly subtract all stellar contributions. Following the technique of BKF, we separate the stellar and AGN contribution from the gaseous emission. Most bright LMXBs are detected in *Chandra* observations and excluded. We then fit the remaining diffuse emission in 0.3-5 keV with a 4-component model, which consists of thermal plasma APEC for hot gas + 7 keV thermal Bremsstrahlung for undetected LMXBs + additional two components (APEC + power-law) for a population of ABs+CVs. In all cases, the hard component for undetected LMXBs is consistent within the error with those expected by extrapolating the X-ray luminosity function of LMXBs (Kim & Fabbiano 2010; see also section 3.4.2 in BKF). For ABs and CVs, we use the spectral parameters determined with the *Chandra* spectra of M31 and M32 where all LMXBs can be detected and removed (see Appendix in BKF). The normalizations are scaled, based on the K-band luminosity within the region of interest. Although the error of the AB+CV component is not reflected in the statistical error of $L_{X,Gas}$, its contribution to the error of $L_{X,Gas}$ is small. For extremely gas-poor galaxies where $L_{X,Gas}$ is lower than that of the soft (APEC) component of AB+CV, we consider the error of the soft component of the AB+CV components (taken from BKF) and add the corresponding error in quadrature to the statistical error. The error of the hard component (power-law) of AB+CV may affect $L_{X,LMXB}$, but its effect is negligible. We note that our measurements are somewhat improved over those of BKF by applying Cash statistics for low s/n data and/or by grouping with a large number of counts, particularly when a large portion of hot gas emission is embedded with brighter point sources (e.g., in NGC 821).

Table 2. Hot gas properties measured in this work

| name | obsid | exp (ksec) | R (") | T (keV) | | | $L_{X,Total}$ ($10^{40}$ erg/s) | $L_{X,Gas}$ ($10^{40}$ erg/s) | | |
|---|---|---|---|---|---|---|---|---|---|---|
| (1) | (2) | (3) | (4) | (5) | | | (6) | (7) | | |
| N0821 | a | 209 | 30 | 0.09 | ( -- | -- ) | 0.883 | 0.025 | (-0.020 | +0.022) |
| N3377 | 02934 | 39 | 30 | 0.19 | ( -- | +0.07) | 0.306 | 0.010 | (-0.006 | +0.007) |
| N3379 | b | 324 | 90 | 0.25 | (-0.02 | +0.02) | 0.864 | 0.042 | (-0.016 | +0.016) |
| N4494 | 02079 | 15 | 30 | 0.62 | (-0.34 | +0.22) | 1.440 | 0.097 | (-0.077 | +0.078) |
| N4564 | 04008 | 17 | 30 | 0.27 | ( -- | +0.45) | 0.285 | 0.038 | (-0.019 | +0.020) |
| N4697 | c | 132 | 60 | 0.31 | (-0.00 | +0.01) | 1.252 | 0.184 | (-0.019 | +0.038) |

(1) Galaxy name
(2) Chandra observation IDs
   a. 04006, 04408, 05691, 05692, 06310, 06313, 06314
   b. 01587, 07073, 07074, 07075, 07076
   c. 04727, 04728, 04729, 04730
(3) Total Chandra exposure time in ksec, after excluding background flares
(4) Radius within which the hot gas emission is extracted
(5) Temperature of hot gas
(6) Total X-ray luminosity in 0.3 – 8 keV
(7) X-ray luminosity from the hot gas in 0.3 – 8 keV

For the remaining well-studied gas-rich galaxies, we adopt $L_{X,Gas}$ from the literature. For NGC 5128 (Cen A), which is rather complex with jets and various features, we take $L_{X,Gas}$ (corrected for a different distance and energy band) from Kraft et al. (2003), who analyzed *Chandra* ACIS-I and *XMM-Newton* data. For seven gas-rich Es, we take $L_{X,Total}$ from the ROSAT measurements by O'Sullivan et al. (2001). We correct it for the different distance and subtract $L_{X,LMXB}$ by applying the scaling relation between $L_{X,LMXB}$ and $L_K$ from BKF ($L_{X,LMXB}/L_K = 10^{29}$ erg s$^{-1}$ $L_{K\odot}^{-1}$). Because of the scatter in this relation, $L_{X,LMXB}$ may vary by ~50% (BKF). For N4374 (M84) which has the lowest $L_X$ among these seven galaxies, this scatter could cause an error of 18% in $L_{X,Gas}$. For the other gas-rich galaxies, this error is less than 10%. We added this error in Table 1.

Although derived physical quantities (such as spectral parameters and their spatial variations) are best measured with the more sensitive Chandra data, the total gas luminosity measured by ROSAT data is still robust for most gas-rich galaxies, where some emission from the outskirt may be missed with Chandra's smaller field of view. Because of the limited field of view of the Chandra ACIS chip, the extended gas emission (r > 4') of gas-rich galaxies falls beyond the main ACIS back-illuminated S3 chip and a part of emission also falls in the gaps between S3 and S2 chips (because a target is usually located on axis, which is ~2' off from the center of S3 toward S2). Taking this limitation in mind, we re-measure $L_{X,Gas}$ of 3 gas-rich galaxies (NGC 1407, NGC 4374, and NGC 5846) with smallest angular extents (but still more extended than the S3 field of view) by analyzing S3 and S2 data and confirm that there is no systematic bias caused by using the ROSAT measurements. Our measurements of $L_{X,Gas}$ are about 10-20% lower than those of in Table 1, as expected by the 'missed' emission. For M87, the AGN and jets may not be fully separated in the ROSAT data, but their contributions are only 0.6% (Pellegrini 2010) and 2% (Harris & Krawczynski 2006), respectively, and therefore have no effect on our results.

## 3. THE $L_{X,Gas}$ – $M_{Total}$ RELATION

We show in Figure 1 the $L_{X,Gas}$ - $L_K$ relation for our 14 ETGs. Although the sample is small, Fig. 1 clearly shows that the $L_{X,Gas}$ – $M_{Total}$ relation is tight. The best fit relation in the form of $L_{X,Gas} \sim M_{Total}^\alpha$ has a slope of $\alpha = 2.7 \pm 0.3$ (dashed line in Figure 1). The rms deviation from this best fit is 0.5 dex (or a factor of 3), which is considerably lower than the factor of ~$10^2$ scatter often seen in previous relations between $L_{X,Total}$ and $L_B$ (e.g., Fabbiano 1989, Eskridge et al. 1995a; Ellis et al 2006) and the factor of ~$10^3$ scatter in the $L_{X,Gas}$ – $L_K$ relation (BKF; see below, Figure 2). Even more striking, this relation is extremely tight among the gas-rich ETGs with $L_{X,Gas} > 10^{40}$ erg s$^{-1}$. The only exception is M84 (NGC 4374). Because M84 is known to suffer from on-going ram-pressure stripping (e.g., Randall et al. 2008), the lower $L_{X,Gas}$ (an order of magnitude below from those of other galaxies with similar $M_{Total}$) can be understood. The solid diagonal line indicates the best-fit relation among gas-rich ETGs (excluding M84) with a slope of $\alpha = 3.3 \pm 0.3$. The rms deviation from this best fit is reduced to only 0.128 dex (or a factor of 1.3). This is the tightest relation ever reported in any relation involving the X-ray luminosity of ETGs. If we simplify the relation by fixing $\alpha = 3$ (see Section 4 for its justification), the best fit relation is

$$(L_{X,Gas} / 10^{40} \text{ erg s}^{-1}) = (M_{Total} / 3.2 \times 10^{11} M_\odot)^3 .$$

M87 is the highest in both $L_{X,Gas}$ and $M_{Total}$, therefore this single galaxy may have a strong leverage in dictating the relation. Moreover, because M87 is in the center of the Virgo cluster, its parameters may reflect an entire cluster rather than a single galaxy (although it was already noted with *Einstein* data that the gas in M87 is cooler, less luminous, and less extended than more regular clusters. e.g., section 5.8 in Sarazin 1988). However, if we exclude M87, the relation remains identical, although with a slightly larger error ($\alpha = 3.3 \pm 0.5$) and a slightly larger rms deviation (0.138 dex or a factor of 1.4).

For comparison, we also plot the $L_{X,Gas}$ - $L_K$ relation in Figure 2. In addition to the 14 ETGs of Figure 1, we show other normal ETGs from the BKF sample (smaller, open symbols) which nicely fill the parameter space with intermediate $L_{X,Gas}$. The $L_{X,Gas}$ - $L_K$ relation is very steep, $L_{X,Gas} \sim L_K^{4.5 \pm 0.8}$. If M87 is excluded, the relation becomes slightly flatter ($L_{X,Gas} \sim L_K^{4.0 \pm 0.7}$), but statistically the slope remains the same. This relation can be compared with the widely used $L_{X,Total}$ - $L_B$ relation (e.g., see Figure 48 in Kormendy et al. 2009), but becomes steeper because of the considerably lower $L_{X,Gas}$ compared to the $L_{X,Total}$ in gas-poor galaxies. The solid diagonal line in Figure 2 indicates the expected X-ray luminosity from the population of LMXBs (using the linear relation of $L_{X,LMXB} / L_K = 10^{29}$ erg s$^{-1}$ $L_{K\odot}^{-1}$ taken from BKF). For a large number of ETGs, the hot gas luminosity is lower than the integrated contribution of LMXBs. In extreme cases, the gas luminosity is even lower than that of ABs and CVs, which is about an order of magnitude lower than that of LMXBs (BKF). We note that our $L_{X,Gas}$ - $L_K$ relation is steeper than any previous relation of $L_{X,Total}$ – $L_K$ and also $L_{X,Gsa}$ – $L_K$, if the latter did not fully consider the contribution from ABs and CVs. Consequently, the $L_{X,Gas}$ - $L_K$ relation is much steeper than the $L_{X,Gas}$ - $M_{Total}$ relation and with a considerably larger scatter. The very tight $L_{X,Gas}$ – $M_{Total}$ relation seen among the gas-rich ETGs with $L_{X,Gas} > 10^{40}$ erg s$^{-1}$ disappears in the $L_{X,Gas}$ - $L_K$ relation. Instead, a large range in $L_{X,Gas}$ (a factor of ~$10^3$) is clearly visible among galaxies with similar $L_K = 1 - 2 \times 10^{11}$ $L_{K\odot}$.

## 4. DISCUSSION

As discussed above (Section 1), the large scatter in $L_X/L_B$ (a factor of $10^2$) has been one of the long-standing puzzles in the field of extra-galactic X-ray astronomy, ever since the *Einstein Observatory* provided the first X-ray images of ETG galaxies (see review in Fabbiano 1989). Now, thanks to the *Chandra X-ray Observatory* and new developments in optical observations, we can revisit this relation, exploring the correlation of physically motivated quantities: the amount of hot gas and the independently determined gravitational potential depth of ETGs. As shown in Section 3, we have found a tight correlation between $L_{X,Gas}$ and $M_{Total}$ with a small rms deviation of a factor of 3 in our 14 galaxy sample, which spans the entire range of measurable $L_{X,Gas}$; the rms deviation is even smaller, a factor of 1.3 for the 7 gas-rich galaxies ($L_{X,Gas} > 10^{40}$ erg s$^{-1}$). Mathews et al. (2006) reached a qualitatively similar conclusion, but they have instead used X-ray determined total mass among group-centered elliptical galaxies with $L_{X,Gas} = 10^{41} - 10^{44}$ erg s$^{-1}$, resulting in a larger scatter.

Since the gas temperature reflects the energy input and the depth of the potential well, the $L_{X,Gas}$ –$T_{Gas}$ relation provides a complementary scaling relation to $L_{X,Gas}$–$M_{Total}$. Interestingly, the functional form of this relation, which we have found in Section 3 (power-law with $\alpha = 2.7 \pm 0.3$ or $\alpha = 3.3 \pm 0.3$ for gas-rich ETGs only is consistent to what would be expected from the steep $L_{X,Gas}$-$T_{Gas}$ relation found in BKF ($L_{X,Gas} \sim T_{Gas}^{4..5 \pm 0.6}$, where typically kT = 0.3 -1 keV; see Fig 7-8 in BKF), for a gas in equilibrium. Given that $M_{Total} \sim T_{Gas}^{3/2}$ (virial theorem), we expect $L_{X,Gas} \sim M_{Total}^3$.

Our results indicate that the *total* mass of an ETG is the primary factor in regulating the amount of hot gas retained by the galaxy in the range of $L_{X,Gas} = 10^{38} - 10^{43}$ erg s$^{-1}$ or $M_{Total}$ = a few x $10^{10}$ - a few x $10^{12}$ $M_\odot$. By contrast, we note that the central binding energy (represented by the central stellar velocity dispersion σ) is less important, as shown by the large scatter in $L_{X,Gas}$ – σ relation (see Figure 5 in BKF). As suggested by recent observations, radio-mode AGN feedback appears to provide enough energy to prevent cooling of the hot ISM (e.g., Nulsen 2007; Diehl & Statler 2008; also Fabian 2012). The tight $L_{X,Gas}$-$M_{Total}$ relation implies that, at the present epoch, non-gravitational energy input is less important than the total mass in determining the gas retention capability of ETGs. The energy feedback may scale with the halo mass, if the DM halo determines the super-massive black-hole mass, as suggested by e.g., Booth & Schaye (2010) as a possible variation of the popular relation between the black-hole and bulge masses (however, see also Kormendy & Bender (2011) who pointed out that $M_{BH}$ is not correlated directly with the dark matter halo particularly for bulgeless galaxies). Even if it scales with the dark matter halo mass, feedback, although necessary to prevent cooling, cannot be more important than the halo mass in determining the gas retention capability.

The $L_{X,Gas}$ – $M_{Total}$ relation has more scatter for gas-poor galaxies with $L_{X,Gas}$ < a few x $10^{39}$ erg s$^{-1}$ (Figure 1). A similar trend (i.e., larger scatter for galaxies with lower $L_X$) is also seen in the $L_{X,Gas}$ - $T_{Gas}$ plot (Fig 7 in BKF). These hot-gas-poor galaxies, where the gas is expected to be in the outflow state (e.g., Ciotti et al. 1996), may have a relatively small amount of DM, and consequently be unable to gravitationally confine the hot gas. With a small amount of DM, the gas is likely to be in the outflow state. In this case, various other factors may become significant, which are minor in gas-rich ETGs where DM dominates. Environmental effects could affect the amount of hot gas as a mechanism to remove hot gas by ram pressure stripping (as already seen in M84), or as a tool to better retain hot gas by adding the external pressure from the hotter ambient ICM (see also Mulchaey & Jeltema 2010). However, even removing environmental effects and considering only isolated galaxies, the scatter may persist (see Memola et al. 2009). The dynamical properties and intrinsic shape of ETGs may also become important. On average, at any fixed optical luminosity, rounder systems show larger $L_X$ than flatter galaxies (e.g., Eskridge et al. 1995b). However, flatter systems also possess, on average, higher rotation levels (Sarzi et al. 2013) so that the binding energy is effectively lower (see Ciotti et al. 1996, Pellegrini 1997). Other possibly important effects include the presence of cold ISM (e.g., Li et al. 2011) and rejuvenation (or stellar age), which could increase the stellar feedback (e.g., Sansom et al. 2006).

The $L_{X,Gas}$ - $L_K$ relation (Figure 2) is very steep ($L_{X,Gas} \sim L_K^{4.5 \pm 0.8}$). To better understand this steep relation, we can consider another scaling relation between the total halo mass and the stellar mass. Numerical simulations suggest that the halo mass and the stellar mass (e.g., from SDSS) are related in the mass range of our interest ($L_K > 10^{10.5}$ L$\odot$), following a relation of the form $M_{Total} \sim M_\star^{1.7-2.5}$ at z=0 (e.g., Moster et al. 2013). We find a similar relation between $M_{Total}$ and $L_K$ among our 14 ETGs. If we adopt this relation and assume $M_\star \sim L_K$, given the observed $L_{X,Gas} \sim M_{Total}^3$ relation, we obtain an extremely steep $L_{X,Gas} \sim L_K^{5-7.5}$ relation, which is steeper than but still consistent with the observed relation ($L_{X,Gas}$ - $L_K^{4.5 \pm 0.8}$). We note that even if the $L_{X,Gas}$ - $L_K$ relation is consistent with the other tighter scaling relations, the large scatter in this relation makes it less useful as a predictive tool. For example, one should not use this relation to predict $L_{X,Gas}$ from $L_K$. Mulchaey & Jeltema (2010) reported a similarly steep relation $L_{X,Gas} \sim L_K^{3.9 \pm 0.4}$ among field early-type galaxies, where the environmental effect (ram-pressure stripping, infall etc.) is minimal. It is important to confirm their results after properly subtracting the ABs and CVs emission.

For galaxies not dominated by an AGN, and ignoring the small AB+CV contribution (~10% of $L_{X,LMXB}$) the total X-ray luminosity of an ETG can be written as

$$(L_{X,Total} / 10^{40} \text{ erg s}^{-1}) = (M_{Total} / 3.2 \times 10^{11} M\odot)^3 + (L_K / 10^{11} L_K\odot),$$

Since the stellar mass to light ratio, $M_\star/L_K$, is close to 1 in solar units (Bell et al. 2003) and $M_{Total} / M_\star \sim 2$ for $L_K \sim 10^{11}$ L$_K\odot$ (Deason et al. 2012), the contributions from the hot gas and LMXBs are approximately comparable at $L_K \sim 10^{11}$ L$_K\odot$ or $M_{Total} \sim$ a few x $10^{11}$ M$\odot$. Above this critical $L_X \sim 10^{40}$ erg s$^{-1}$, the hot gas will dominate the total X-ray emission, below it the LMXBs will dominate. This critical $L_X$ is approximately where the hot gas states changes between inflows and outflows. It will be very useful to observationally determine the exact location of the division between inflows and outflows in terms of $M_{Total}$ or $L_{X,Gas}$ to place strong constraints on theoretical model parameters. The currently available sample, however, lacks galaxies with an intermediate mass and hot gas luminosity, in the critical range ($L_{X,Gas} \sim 10^{40}$ erg s$^{-1}$ or $M_{Total} \sim 5 \times 10^{11}$ M$_\odot$).

## 5. SUMMARY AND CONCLUSIONS

In summary, our scaling relations can be written in a simplified form:

$$(L_{X,Gas} / 10^{40} \text{ erg s}^{-1}) = (M_{Total} / 3.2 \times 10^{11} M_\odot)^3 \quad \text{from this work,}$$
$$L_{X,Gas} \sim T_{Gas}^{4.5} \quad \text{from BKF,}$$
$$L_{X,Gas} \sim L_K^{4.5 \pm 0.8} \quad \text{(with a large scatter).}$$

By comparison, the scaling relations for clusters of galaxies, where the gas is hotter at kT = 2–10 keV, are $L_X \sim M^2$ and $L_X \sim T^3$ (e.g., Pratt et al. 2009) with a tendency that the $L_X$-T relation steepens for lower T (< 3 keV) groups (e.g., Eckmiller et al. 2011). It is well known that these relations in clusters are significantly steeper than self-similar expectations, predicting $L_X \sim M^{4/3}$ and $L_X \sim T^2$ (e.g., Eke, Navarro & Frenk 1998; Arnaud & Evrard 1999). Our relations in ETGs are even steeper than those of clusters. The cause of the steeper relations will be addressed in a forthcoming paper (Kim et al. in prep.).

If the presence of a well defined $L_{X,Gas}$-$M_{Total}$ relation with little scatter is confirmed in a larger sample of galaxies, this scaling relation will provide the basis for a new reliable way for measuring the total mass (and in particular DM content) of ETGs.

**Acknowledgement**

We thank Alis Deason, Silvia Pellegrini, Aaron Romanowsky, and Mark Sarzi for helpful discussions. The data analysis was supported by the CXC CIAO software and CALDB. We have used the NASA NED and ADS facilities, and have extracted archival data from the Chandra archives. This work was supported by NASA contract NAS8-03060 (CXC).

References

Arnaud, M. & Evrard, A. E. 1999, MNRAS, 305, 631
Bender, R., Surma, P., Dobereiner, S., Mollenhoff, C., & Madejsky, R. 1989, A&A, 217, 35
Binney, J. 2004, MN, 347, 1093
Booth, C. M. & Schaye, J. 2010, MNRAS Lett. 405, L1
Boroson, B, Kim, D.-W., and Fabbiano, G. 2011, ApJ, 729, 12 (BKF)
Canizares, C. R., Fabbiano, G., & Trinchieri, G. 1987, ApJ, 312, 503
Cappellari, M. et al. 2011, MNRAS, 413, 813
Ciotti, L., D'Ercole, A., Pellegrini, S., & Renzini, A. 1991, ApJ, 376, 380
Ciotti, L., Pellegrini, S., 1996, MNRAS 279, 240
David, L. P., Jones, C., & Forman, W. 1991, ApJ, 369, 121
Deason et al. 2012, ApJ, 748, 2
Diehl, S., & Statler, T. 2008, ApJ, 680, 897
Eckmiller, H. J., Hudson, D. S. & Reiprich, T. H. 2011, A&A, 535, 105
Eke, V. R., Navarro, J. F. & Frenk, C. S. 1998, ApJ, 503, 569
Ellis, S. C., & O'Sullivan, E. 2006, MNRAS, 367, 627
Eskridge, P., Fabbiano, G., & Kim, D.-W. 1995a, ApJS, 97, 141
Eskridge, P., Fabbiano, G., & Kim, D.-W. 1995b, ApJ, 442, 523
Fabbiano, G. 1989, ARA&A, 27, 87
Fabbiano, G. 2006, ARA&A, 44, 323
Fabian, A. C. 2012 ARAA, 50, 455
Forman, W., Jones, C., & Tucker, W. 1985, ApJ, 293, 102
Harris, D. E. &, H., 2006, ARAA 44, 463
Kim, D.-W., & Fabbiano, G. 2010, ApJ, 721, 1523
Kormendy, J. et al. 2009, 182, 216
Kormendy, J. & Bender, R. 2011, *Nature*, 469, 377–80
Kraft, R. P. et al. 2003, ApJ, 592, 129
Li, J.-T., et al. 2011 ApJ, 737, 41
Mathews, W. & Brighenti, F. 2003, ARAA, 41, 191
Mathews, W., et al. 2006, ApJ, 652, L17
Moster et al. 2013, MNRAS, 428, 3121
Mulchaey, J. S. & Jeltema, T. E. 2010, ApJ, 715, L1
Nipoti, & Binney, J. 2007, MN, 382, 1481
Nulsen, P., et al. 2007 ESO Astrophys. Symp., ed. Boehringer, p210 Springer-Verlag
O'Sullivan, et al. 2001, MNRAS, 328, 461
Pellegrini, S. 2005, MNRAS, 364, 169
Pellegrini, S. 2010, ApJ, 717, 640
Pellegrini, S., Held, E. V., & Ciotti, L. 1997, MNRAS, 288, 1
Randall, S. et al. 2008, ApJ, 688, 208
Sansom, A. E., et al. 2006, MN, 370, 1541
Sarazin, C. L. 1988 "X-ray emissions from clusters of galaxies" Cambridge Univ. Press
Sarzi, M et al. 2013, arXiv/1301.2589
Tonry, J. L. et al. 2001, ApJ, 546, 681
Trinchieri, G., & Fabbiano, G. 1985, ApJ, 269, 447
White III., R. E., & Sarazin, C. L. 1991, ApJ, 367, 476

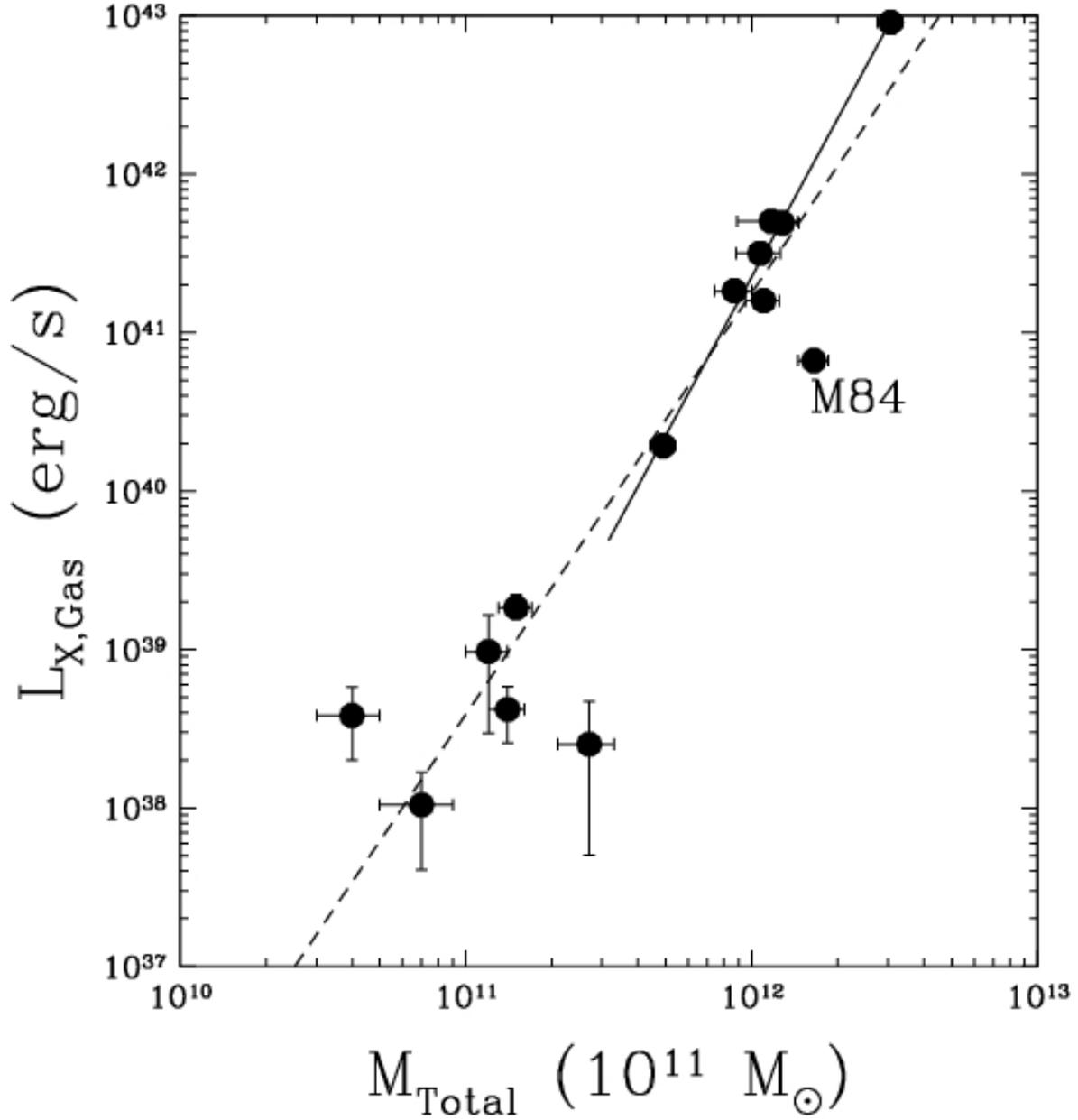

**Fig 1.** X-ray luminosities from the hot gas are plotted against $M_{Total}$ (< 5 $R_e$) taken from Deason et al. (2012). The dashed line indicates the best fit using the entire sample ($L_{X,Gas} \sim M_{total}^{2.7}$) and the solid line is for gas-rich ETGs with $L_{X,Gas} > 10^{40}$ erg s$^{-1}$ ($L_{X,Gas} \sim M_{total}^{3.3}$).

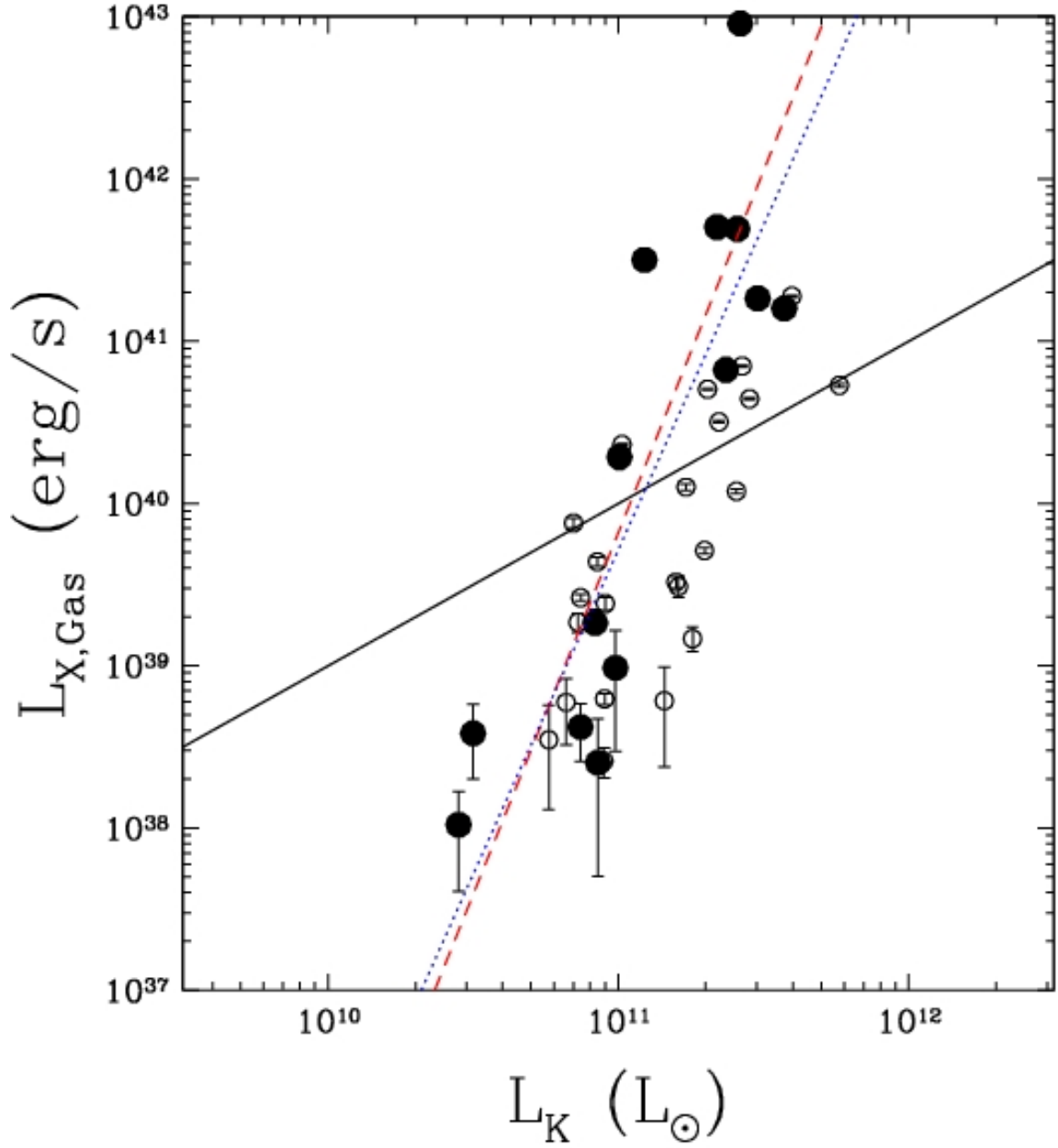

**Fig 2.** X-ray luminosities from the hot gas are plotted against $L_K$. The small open circles are additional galaxies from the BKF sample. The solid line indicates $L_X$ expected from LMXBs. The (red) dashed and (blue) dotted lines indicate the best fit relations with ($L_{X,Gas} \sim L_K^{4.5}$) and without M87 ($L_{X,Gas} \sim L_K^{4}$).